\documentclass[sigconf]{acmart}
\usepackage{graphicx}
\usepackage{color}
\usepackage{xspace}
\usepackage{amsmath,amsfonts}
\usepackage{algorithm}
\usepackage{algorithmic}
\usepackage{textcomp}
\usepackage{xcolor}
\usepackage{booktabs}
\usepackage{booktabs,xltabular}
\usepackage{hyperref}
\usepackage{hyperxmp}
\usepackage{wrapfig}
\usepackage{float} 
\usepackage{caption} 
\usepackage{float}
\usepackage{tcolorbox}
\usepackage{listings}
\usepackage{multirow}

\newcommand{\squishlist}{
 \begin{list}{$\bullet$}
   { \setlength{\itemsep}{0pt}
     \setlength{\parsep}{0pt}
     \setlength{\topsep}{0pt}
     \setlength{\partopsep}{0pt}
     \setlength{\leftmargin}{2.5em}
     \setlength{\labelwidth}{1.5em}
     \setlength{\labelsep}{0.5em} } }

\def\BibTeX{{\rm B\kern-.05em{\sc i\kern-.025em b}\kern-.08em
    T\kern-.1667em\lower.7ex\hbox{E}\kern-.125emX}}

\newcommand{\squishend}{
  \end{list}  }

\newcommand{\TheName}{K-ASTRO}

\begin{document}

\title{\TheName{}: Structure-Aware Adaptation of LLMs for Code Vulnerability Detection\\
}


\author{Yifan Zhang*}\thanks{*Equal contribution; authors listed in alphabetical order.}
\affiliation{%
  \institution{Vanderbilt University}
  \city{Nashville}
  \state{Tennessee}
  \country{USA}
}

\author{Michael Sandborn*}
\affiliation{%
  \institution{Vanderbilt University}
  \city{Nashville}
  \state{Tennessee}
  \country{USA}
}

\author{Stefan Larson}
\affiliation{%
  \institution{Vanderbilt University}
  \city{Nashville}
  \state{Tennessee}
  \country{USA}
}

\author{Yu Huang}
\affiliation{%
  \institution{Vanderbilt University}
  \city{Nashville}
  \state{Tennessee}
  \country{USA}
}

\author{Kevin Leach}
\affiliation{%
  \institution{Vanderbilt University}
  \city{Nashville}
  \state{Tennessee}
  \country{USA}
}


\begin{abstract}

Large Language Models (LLMs) are transforming software engineering tasks, including code vulnerability detection—a critical area of software security. However, existing methods often rely on resource-intensive models or graph-based techniques, limiting their accessibility and practicality. This paper introduces \TheName{}, a lightweight Transformer model that combines semantic embeddings from LLMs with structural features of Abstract Syntax Trees (ASTs) to improve both efficiency and accuracy in code vulnerability detection. Our approach introduces an AST-based augmentation technique inspired by mutation testing, a structure-aware attention mechanism that incorporates augmented AST features, and a joint adaptation pipeline to unify code semantics and syntax. Experimental results on three large-scale datasets—BigVul, DiverseVul, and PrimeVul—demonstrate state-of-the-art performance while enabling rapid inference on CPUs with minimal training time. By offering a scalable, interpretable, and efficient solution, \TheName{} bridges the gap between LLM advancements and practical software vulnerability detection, providing open-sourced tools to foster further research.

\end{abstract}

\maketitle

\section{Introduction}\label{sec:intro}

Large Language Models (LLMs) have demonstrated remarkable capabilities in tasks such as question answering, code generation, and text summarization~\cite{kamalloo-etal-2023-evaluating, zan-etal-2023-large-nl2code, zhang-etal-2023-extractive-summarization}. Built upon the Transformer architecture~\cite{vaswani2017attention}, LLMs leverage large datasets to solve domain-specific problems with unprecedented efficiency, making them increasingly integral in software engineering. Among the various applications of LLMs, code vulnerability detection holds particular significance, where the goal is to determine whether a given piece of code is vulnerable to security threats. Early and reliable detection of vulnerabilities minimizes the risk of exploitation and reduces the cost of addressing these issues later in the software lifecycle.

Effectively leveraging LLMs for vulnerability detection presents two key challenges. First, pre-training or fine-tuning Transformer-based LLMs is computationally demanding, often requiring extensive GPU resources unavailable to many practitioners~\cite{zhang2023ecoassistant}. Second, capturing the intricate nuances of software vulnerabilities in standalone code functions remains a significant hurdle. For example, it is challenging to infer whether inputs to a function have been sanitized prior to their use. While pretrained models~\cite{feng2020codebert, wang2021codet5, guo2022unixcoder, touvron2023llama} with hundreds of millions to billions of parameters offer state-of-the-art baselines, they remain resource-intensive and underperform on nuanced tasks like vulnerability detection when used off-the-shelf~\cite{chatgpt-eval-vd-cheshkov-2023, purba-2023-llm-svd, chatgpt-for-vd-2023-fu, ding2023traced}. Notably, GitHub employs LLMs with CodeQL~\cite{codeql}, combining generative models with static code analysis to identify vulnerabilities at scale, but this setup requires significant infrastructure investments.

In this paper, we introduce \TheName{}, a novel and lightweight Transformer-based model that combines semantic embeddings from LLMs with structural features derived from Abstract Syntax Trees (ASTs) to enhance code vulnerability detection. Our approach addresses the challenges of efficiency and nuance by introducing three key innovations: (i) \textbf{Diversity-Introducing AST Augmentation}, which enhances feature diversity through an AST-based mutation method inspired by mutation testing; (ii) \textbf{Structure-Aware Attention Bias}, a novel mechanism that incorporates augmented AST features into the Transformer block, guiding the model’s attention to structural relationships; and (iii) \textbf{Joint LLM Adaptation}, a training pipeline that unifies structural and semantic information for improved prediction accuracy. These innovations bridge the gap between off-the-shelf LLM capabilities and the domain-specific requirements of code vulnerability analysis. Our contributions are as follows:

\begin{itemize}
    \item We propose \TheName{}, a lightweight, single-layer, encoder-only Transformer that unifies code syntax (via AST features) and semantics (via LLM embeddings), significantly improving both binary vulnerability prediction and CWE classification.
    \item We evaluate \TheName{} on three large-scale, real-world datasets: BigVul, DiverseVul, and PrimeVul. These datasets cover hundreds of open-source projects. \TheName{} achieves state-of-the-art performance with minimal computational requirements.
    \item We conduct an ablation study to validate \TheName{}’s design, comparing it with simpler models that classify code embeddings without leveraging AST structure.
    \item To foster further research, we open-source all code, datasets, and tools, including scripts for data preprocessing, LLM API interactions, model training, evaluation, and embedding generation.
\end{itemize}

The remainder of this paper is organized as follows: Section~\ref{sec:problem} outlines the problem formulation and background. Section~\ref{sec:approach} describes the datasets and architectural details of \TheName{}. Section~\ref{sec:experiments} presents experimental results, guided by four research questions. Section~\ref{sec:related} discusses related work in LLM-based vulnerability detection. Finally, Section~\ref{sec:conclusion} summarizes our findings and highlights the limitations of our approach.

\section{Problem Statement}\label{sec:problem}

In this section, we introduce the problem of vulnerability detection, motivate our approach with an example of vulnerable code, and describe the Abstract Syntax Tree (AST) representation that underpins our design.

\subsection{Vulnerability Detection}

Detecting vulnerabilities in real-world software is a challenging task, particularly given the sheer scale of modern codebases—often comprising millions of lines of code contributed by multiple developers. Despite rigorous testing~\cite{ammann2016introduction}, vulnerabilities persist due to their dispersed nature and the potential for fixes to inadvertently introduce new issues. Common automated approaches, such as static analysis~\cite{zheng2006value}, fuzzing~\cite{zhu2022fuzzing}, and software testing~\cite{wu2023automated}, often suffer from high rates of false positives and negatives.

Vulnerability detection can be categorized into two complementary tasks: (i) \textit{vulnerability prediction}, which determines \textit{whether} a piece of code is vulnerable, and (ii) \textit{CWE classification}, which identifies \textit{how} the code is vulnerable by mapping it to a Common Weakness Enumeration (CWE). CWE identifiers provide an abstract description of vulnerability types, while specific instances of vulnerabilities are documented as Common Vulnerabilities and Exposures (CVEs). Our approach addresses both tasks by leveraging the structural and semantic information encapsulated in Abstract Syntax Trees (ASTs).

\subsection{Motivating Example}

Code vulnerabilities often arise from issues such as buffer overflows or improper memory management. Consider the example in Listing~\ref{lst:vuln_code_sample}, which lacks null termination in its arrays and uses an insecure call to \texttt{strcpy()}. This omission risks injecting unexpected data, making the code vulnerable. For binary vulnerability prediction, this function would be labeled as \texttt{1} (vulnerable), and for CWE classification, it would be assigned CWE-170, ``Improper Null Termination''~\footnote{\url{https://cwe.mitre.org/data/definitions/170.html}}. Our methodology uses the AST of the source code to effectively represent and analyze such vulnerabilities, as described further in Section~\ref{ast_rep}.

\begin{figure}[ht]
\centering
\begin{lstlisting}[language=C, caption={\textbf{Example Vulnerable Function.} Vulnerable C code that lacks null termination, resulting in a risky \texttt{strcpy()} call.}, label={lst:vuln_code_sample}]
#include <stdio.h>
#include <string.h>
#include <unistd.h>
#define MAXLEN 1024

int main() {
    char *inputbuf;
    char pathbuf[MAXLEN];
    read(0, inputbuf, MAXLEN);
    strcpy(pathbuf, inputbuf);
    return 0;
}
\end{lstlisting}
\end{figure}

\begin{figure}
\centering
    \includegraphics[width=0.5\textwidth]{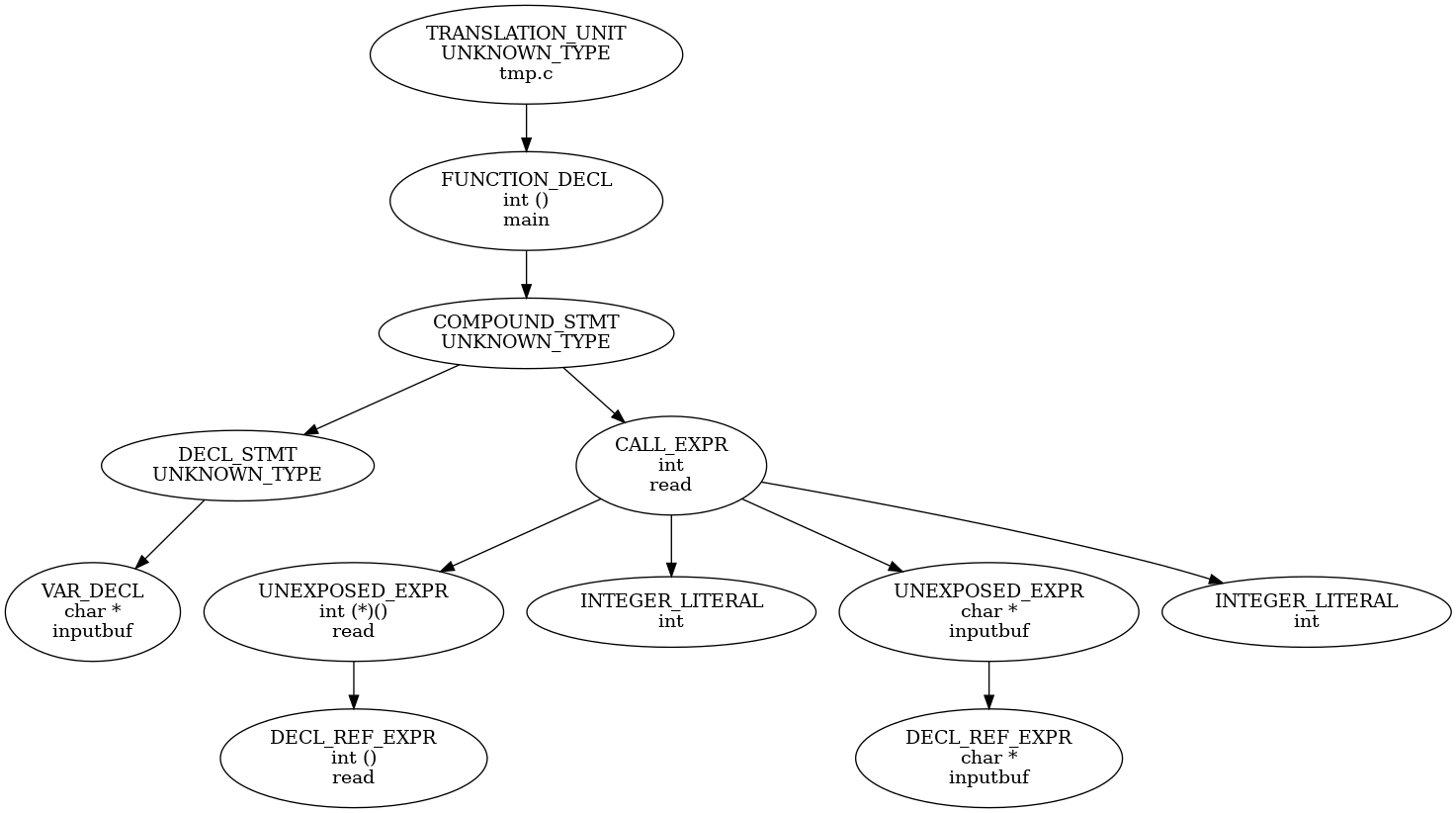}
    \caption{\textbf{Example AST.} AST representation of lines 6, 7, and 9 of Listing~\ref{lst:vuln_code_sample}, parsed with Clang 14.0 and visualized with Graphviz.}
    \label{fig:vuln_code_sample_ast}
\end{figure}

\subsection{Abstract Syntax Tree Representation}
\label{ast_rep}

The Abstract Syntax Tree (AST) is a hierarchical representation of a program's structure, where each node corresponds to a construct in the source code. ASTs are generated during the parsing phase of compilation and encode both syntactic and semantic relationships, such as variable declarations, control flow, and data dependencies. Internal nodes represent operators, while leaf nodes represent operands, and edges capture relationships such as loop conditions and variable assignments.

This structured representation is widely used in machine learning models for tasks like code classification~\cite{wang_code_clones, wang_ast_xlang_cls, icse_novel_ast_code_rep}. In our approach, we enhance the standard AST by introducing augmented variants. Specifically, we replace selected nodes with subtrees from other functions sharing the same vulnerability label, rooted at matching node types (Section~\ref{sec:ast_augmentation}). Figure~\ref{fig:vuln_code_sample_ast} shows part of the AST derived from Listing~\ref{lst:vuln_code_sample}. This augmentation improves the model's ability to generalize across diverse code samples while maintaining the structural context essential for accurate vulnerability detection.

\section{Approach: K-ASTRO}\label{sec:approach}

We address the related tasks of vulnerability prediction (binary classification) and CWE classification from source code. Given a standalone source code function \( f \in \mathcal{F} \) in C/C++, the goal is to reliably predict either the presence of a vulnerability or the corresponding Common Weakness Enumeration (CWE) exhibited in the function, without leveraging surrounding code context such as function callers or repository-level information.

To achieve this, we propose \TheName{}, a lightweight yet powerful framework that addresses the limitations of existing approaches by combining structural and semantic information effectively. The framework integrates three core components that work in tandem to improve vulnerability detection accuracy:

\textbf{First,} \textbf{Diversity-Introducing AST Augmentation} enriches the structural representation of code by introducing controlled variations in the Abstract Syntax Tree (AST). This step ensures that the model is exposed to a diverse range of structural patterns during training, making it more robust to real-world scenarios where vulnerable patterns may vary significantly. By enhancing structural diversity, this component helps the model generalize better to unseen data.

\textbf{Second,} \textbf{Structure-Aware Attention Bias} leverages these augmented ASTs to encode structural patterns directly into the Transformer’s attention mechanism. This step injects a deeper understanding of code structure into the model by highlighting the relationships and interactions between nodes in the AST. The integration of co-occurrence patterns allows the attention mechanism to focus on the most relevant structural features, reducing noise and improving the model’s ability to identify vulnerabilities.

\textbf{Finally,} \textbf{Joint LLM Adaptation} brings everything together by combining the semantic information captured by pre-trained LLM embeddings with the structural insights derived from the augmented ASTs. This joint representation bridges the gap between the high-level semantic understanding of code and the low-level structural details, ensuring that the model benefits from both perspectives. Together, these components enable \TheName{} to provide accurate and efficient vulnerability predictions.

\begin{figure}[ttbp]
\centering
\includegraphics[width=\columnwidth]{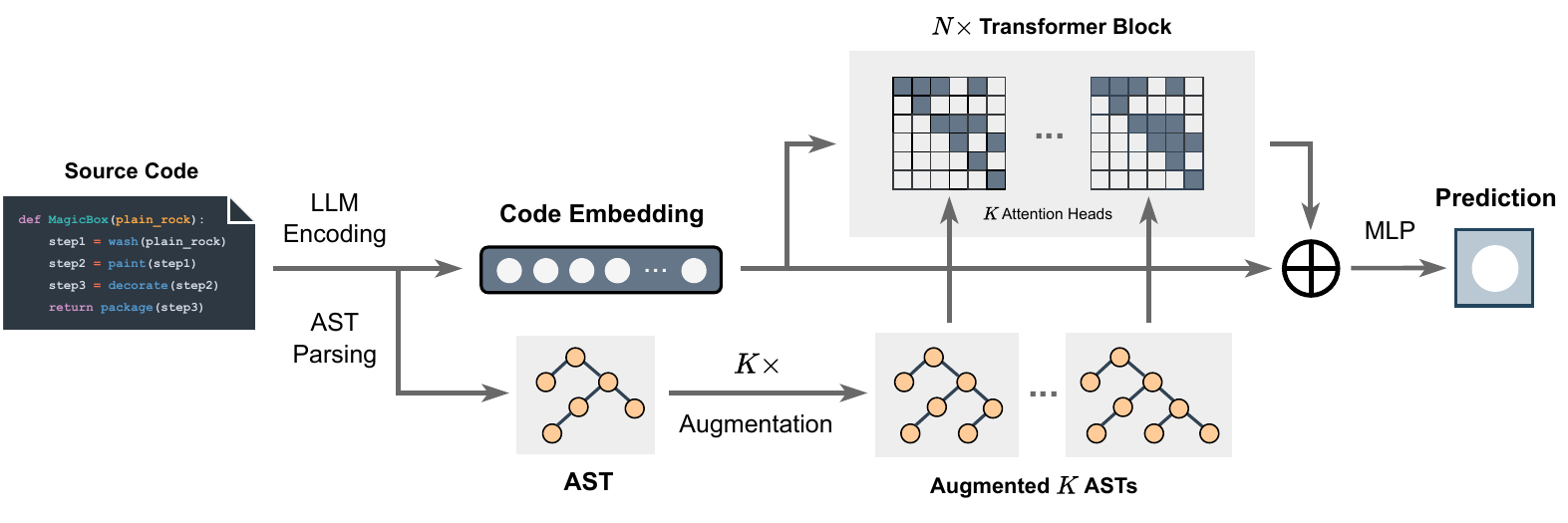}
\caption{\textbf{Overview of K-ASTRO.} The framework processes source code into semantic embeddings via LLMs and structural embeddings via AST augmentation. Augmented ASTs provide structural insights through a structure-aware attention mechanism, which is combined with LLM embeddings for final vulnerability prediction using a single lightweight Transformer block.}
\label{fig:overview}
\end{figure}

\subsection{Diversity-Introducing AST Augmentation}
\label{sec:ast_augmentation}

The first step in \TheName{} is to address the inherent sparsity and diversity of vulnerability patterns in real-world codebases. Vulnerable code fragments often constitute a small fraction of large repositories, making it difficult for models to generalize effectively. To mitigate this issue, we introduce a novel AST augmentation technique inspired by mutation testing.

\textbf{AST Generation and Subtree Extraction:} For a given source code function \( f \), the AST \( T(f) \) is parsed using Clang. Subtrees are extracted recursively via depth-first search and stored in a catalog \( \mathcal{C}_{\text{AST}} \), organized by CWE classes \( c \). This catalog serves as a repository of structural patterns for augmentation.

\textbf{Augmentation via Subtree Replacement:} To augment the AST, we randomly select a node \( v \) in \( T(f) \) and replace the subtree rooted at \( v \) with another subtree \( s' \) from \( \mathcal{C}_{\text{AST}}(c) \), ensuring that the root types match. This process generates a structurally diverse augmented AST \( T_{\text{aug}}(f) \), which is formally expressed as:
\[
T_{\text{aug}}(f) = \mathcal{A}_{\text{aug}}(T(f), \mathcal{C}_{\text{AST}}).
\]

\begin{figure}[ttbp]
\centering
\includegraphics[width=1\linewidth]{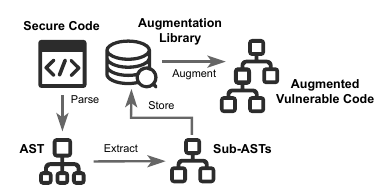}
\caption{\textbf{AST Augmentation Pipeline.} The process involves AST generation, subtree extraction, and augmentation through node replacement to create structurally diverse representations for vulnerability detection.}
\label{fig:ast_lib}
\end{figure}

This augmentation introduces controlled variations that enhance the model’s ability to learn robust representations of vulnerability patterns. While the augmented ASTs may not preserve the exact semantics of the original code, they provide a rich structural feature set that complements the model’s learning process. Figure~\ref{fig:ast_lib} illustrates the augmentation pipeline.

\subsection{Structure-Aware Attention Bias}
\label{sec:structure_aware}

Building on the augmented ASTs, the second component of \TheName{} focuses on integrating structural context into the model’s attention mechanism. While traditional attention mechanisms treat all input tokens equally, our structure-aware attention bias prioritizes important structural relationships derived from the ASTs.

\textbf{Co-Occurrence Matrix Construction:} For each AST \( T(f) \), we compute a co-occurrence matrix \( M \) that captures adjacency relationships between node types:
\[
M_{ij} = \text{frequency}(t_i \to t_j), \quad \forall t_i, t_j \in T(f).
\]

\textbf{Logarithmic Binning and Bias Computation:} To prevent the model from overfitting to specific patterns, we apply logarithmic binning to the co-occurrence values:
\[
B_{ij} = \lceil \log_{10}(M_{ij} + 1) \rceil.
\]

\textbf{Bias Integration into Attention:} For \( K \) AST variants (one original and \( K-1 \) augmented), we compute a separate bias matrix \( B^k \) for each variant and combine them to form the final bias matrix:
\[
B_{\text{max}, ij} = \max(B^1_{ij}, B^2_{ij}, \dots, B^K_{ij}).
\]

The final attention map is computed by adding this bias to the original attention map \( A \):
\[
A' = A + B_{\text{max}}.
\]

\begin{figure}[ttbp]
\centering
\includegraphics[width=1\linewidth]{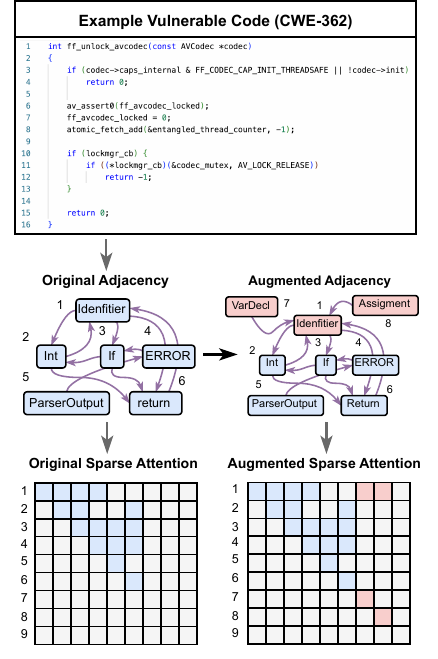}
\caption{\textbf{Structure-Aware Attention Mechanism.} The mechanism incorporates co-occurrence matrices from ASTs into the Transformer’s attention map, enhancing its structural understanding.}
\label{fig:model_detail}
\end{figure}

This structure-aware bias ensures that the attention mechanism focuses on the most relevant structural features, improving the model’s ability to identify vulnerabilities. Figure~\ref{fig:model_detail} provides an overview of this mechanism.

\subsection{Joint LLM Adaptation}
\label{sec:joint_adaptation}

The final component of \TheName{} unifies the structural and semantic embeddings into a joint representation, ensuring that the model benefits from both detailed structural insights and high-level semantic understanding. By combining these complementary perspectives, \TheName{} delivers robust and efficient predictions for both binary vulnerability detection and CWE classification.

The semantic embedding \( T \) is derived from a pre-trained LLM and encodes the high-level contextual meaning of the source code. It represents the function as a dense vector in \( \mathbb{R}^d \), capturing linguistic and semantic nuances. In contrast, the structural embedding \( A' \), generated through the structure-aware attention mechanism, encodes hierarchical relationships and structural interactions within the AST. These two embeddings reflect different but equally important aspects of the source code, making their combination a powerful tool for vulnerability detection.

To integrate these embeddings, we concatenate \( T \) and \( A' \) to form a unified representation:
\[
\text{Input to Classifier: } [A'; T].
\]
This joint representation ensures that the model can simultaneously leverage high-level semantic context and low-level structural details. The combined embedding is passed through a ResNet-styled MLP classifier, where the concatenated vector undergoes transformation through learnable parameters:
\[
L = \sigma(W \cdot [A'; T] + b),
\]
where \( W \) is the weight matrix, \( b \) is the bias term, and \( \sigma \) is an activation function such as ReLU. This architecture ensures stable gradient flow during training and provides sufficient capacity for learning complex patterns inherent in vulnerabilities.

By unifying the semantic and structural embeddings, \TheName{} addresses both the broad contextual variability in codebases and the localized, nuanced patterns of vulnerabilities. This approach not only improves prediction accuracy but also maintains computational efficiency. The joint adaptation mechanism allows \TheName{} to scale effectively to large datasets while remaining lightweight enough for practical deployment, demonstrating its utility in real-world software engineering scenarios.

\section{Experiments}
\label{sec:experiments}

In this section, we describe the datasets considered in our experiments and the data preparation process. The following four research questions (RQs) guide our investigation:

\begin{enumerate}
    \item \textbf{RQ1: Other LLMs vs. \TheName{}} How well does \TheName{} perform in comparison to prompting common off-the-shelf LLMs for vulnerability prediction and CWE classification, and in comparison to recent larger models fine-tuned on code?
    \item \textbf{RQ2: CWE-Specific Performance} What trends exist in the performance of \TheName{} for specific CWE classes?
    \item \textbf{RQ3: Model Efficiency} Here we assess the training overhead of \TheName{} by considering the number of parameters in the model, training time, and inference throughput on different datasets.
    \item \textbf{RQ4: Ablation Study} We compare \TheName{} to 3 simpler models trained on the same CWE classification for C/C++ source code to justify the proposed \TheName{} model.
\end{enumerate}

\subsection{Datasets and Data Summary}
\label{sec:datasets}

We focus on binary vulnerability classification and multi-class CWE classification of C/C++ source code functions, utilizing three large-scale datasets: BigVul~\cite{bigvul}, DiverseVul~\cite{diversevul}, and PrimeVul~\cite{primevul}. Table~\ref{tab:dataset_summary} provides a summary of the datasets, including train, validation, and test split sizes, as well as the distribution of vulnerable and non-vulnerable samples and the number of unique CWE classes.

\begin{table}[h]
    \centering
    \caption{Summary of datasets used in this study.}
    \label{tab:dataset_summary}
    \resizebox{\columnwidth}{!}{%
    \begin{tabular}{l|c|c|c|c|c|c}
        \toprule
        \textbf{Dataset} & \textbf{Train} & \textbf{Val} & \textbf{Test} & \textbf{Vuln} & \textbf{Not Vuln} & \textbf{CWEs} \\
        \midrule
        BigVul~\cite{bigvul} & 148,067 & 32,045 & 31,978 & 170,613 & 41,477 & 36 \\
        DiverseVul~\cite{diversevul} & 206,962 & 24,615 & 24,813 & 207,632 & 48,758 & 49 \\
        PrimeVul~\cite{primevul} & 183,673 & 25,211 & 25,706 & 83,191 & 151,399 & 4 \\
        \bottomrule
    \end{tabular}%
    }
\end{table}

The datasets are selected to ensure diverse sources and high-quality labeling:
\begin{itemize}
    \item \textbf{BigVul} is derived from CVE database crawls, featuring functions from 91 CWEs with an emphasis on code prior to bug-fixing commits.
    \item \textbf{DiverseVul} introduces a larger variety of CWEs with a more recent collection strategy, focusing on eliminating heuristic biases from commit messages.
    \item \textbf{PrimeVul} enhances dataset quality through rigorous de-duplication and chronological data splitting, addressing data leakage issues inherent in prior datasets.
\end{itemize}

\subsection{Data Preparation}
\label{sec:dataprep}

To ensure consistency and compatibility across the datasets used in this study, we followed a structured data preparation pipeline, which is outlined below.

\textbf{Pre-Processing.} We converted each dataset into Parquet files while preserving train, validation, and test splits. To enable consistent tracking, a unique identifier (UUID) was added to each function. Invalid CWE entries, such as empty strings or lists, were mapped to a "No CWE" class, and rows containing multiple CWEs were discarded. Source code comments were removed using the CodeTF ApexCodeUtility~\cite{salesforcecodetf}. Token counts for each function were calculated using the \texttt{tiktoken}\footnote{\url{https://pypi.org/project/tiktoken/}} library, ensuring compatibility with embedding models.

\textbf{Embedding Collection.} We utilized OpenAI’s \textit{text-embedding-ada-002}\footnote{\url{https://platform.openai.com/docs/guides/embeddings}} and \textit{text-embedding-3-small}\footnote{\url{https://platform.openai.com/docs/api-reference/embeddings}} models to generate 1536-dimensional embeddings for each function. Functions exceeding the token limit of 8191 were excluded. This process produced approximately 600,000 embeddings across all splits, which were used as input representations for classification.

\textbf{AST Collection and Augmentation.} Abstract Syntax Trees (ASTs) for each function were extracted using Clang version 14.0 and stored in JSON format. Subtrees were recursively collected and categorized by CWE label and root node type. For augmentation, we randomly replaced AST nodes with subtrees that matched the function’s original CWE label. This procedure generated $K=4$ augmented samples per function, enriching the training data.

\textbf{Adjacency Matrix Generation.} After generating the augmented ASTs, we vectorize each AST by producing an adjacency matrix based on the node types present in the AST. We cap the maximum number of node types at 64 and observed 50 node types in practice from Clang. The dimensionality of this matrix directly affects the architectural parameters of the \TheName{} model (Section~\ref{sec:approach}). Each of these augmented matrices is incorporated into \TheName{} using sparse attention mechanisms, as detailed in Section~\ref{sec:structure_aware}.

\begin{figure*}
    \centering
    \includegraphics[width=\linewidth]{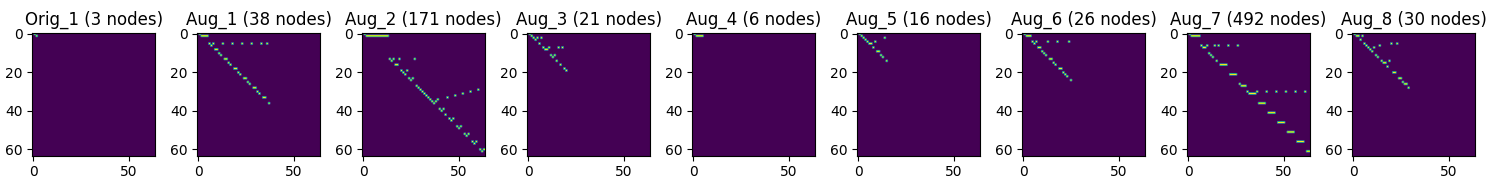}
    \caption{\textbf{Visualization of Augmented ASTs.} Adjacency matrices generated during the AST augmentation process, where subtrees from functions with corresponding vulnerability labels replace selected nodes in the original ASTs. The original matrix (left column) and $K=8$ augmented matrices (right columns) depict node connectivity in the resulting ASTs. We consider a total of $n=50$ node kinds, as observed in the datasets using Clang. These structures are incorporated into \TheName{} via sparse attention to enhance vulnerability prediction performance (Section~\ref{sec:structure_aware}). We set $K=4$ in our experiments to balance feature diversity and experimentation speed.}
    \label{fig:aug_ast_sample}
\end{figure*}

\subsection{Implementation and Training Details}

To train and evaluate \TheName{}, we designed an efficient implementation strategy, ensuring reproducibility and scalability across datasets and experiments.

\textbf{Model Architecture.} \TheName{} integrates an MLP and a Transformer layer in its joint adaptation module. The MLP has a hidden dimension of 512 and 3 layers, while the Transformer is a single-layer encoder with a hidden size of 64. These dimensions balance expressiveness and computational efficiency, enabling effective modeling of source code features and structural representations.

\textbf{Training Configuration.} The model was trained using the Adam optimizer~\cite{kingma2014adam}, with a batch size of 32 and a learning rate of $1\mathrm{e}{-3}$. A total of 25 epochs were conducted, comprising 5 runs of 5 epochs each with distinct random seeds to ensure robust evaluation. To incorporate augmented AST adjacency matrices, we set $K=4$, which balances the diversity of syntactic patterns introduced during training with computational efficiency.

\textbf{Hardware and Efficiency.} Training was performed on a single NVIDIA RTX A6000 GPU with 48GB of VRAM, while inference was tested on both GPU and Intel(R) Xeon(R) Gold 6330N CPUs. Training time ranged from 1 to 3 hours per experiment depending on dataset size, while inference over test sets, containing 24,000 to 32,000 samples, took approximately 10–30 seconds per dataset.

\textbf{Scalability.} Despite its small size of 1 million trainable parameters, \TheName{} demonstrated competitive performance with minimal computational resources. The model’s ability to train and infer efficiently makes it suitable for practical applications in resource-constrained environments.

\subsection{Prompted LLM Performance}
\label{sec:prompted_llm_performance}

To evaluate the feasibility of using general-purpose LLMs for vulnerability classification tasks, we conducted experiments with three popular models: GPT-3.5, GPT-4o, and Claude 3 Haiku. These evaluations focus on assessing their performance on CWE classification and binary vulnerability prediction, comparing them against the specialized model, \TheName{}.

\textbf{Evaluation Setup.} We used programmatic APIs to query these models for predictions, using the BigVul dataset's test set as input. For each sample, we requested outputs in JSON format: 
\texttt{is\_vuln} (binary label indicating vulnerability), 
\texttt{cwe\_label} (CWE class prediction), 
and \texttt{reasoning} (a brief explanation in up to 100 words). Chain-of-thought prompting techniques~\cite{llmcot} were adopted to refine the prompt, incorporating explicit format instructions, a synopsis of vulnerabilities, and examples of CWE categories. Due to malformed JSON outputs from Claude ($\sim$72\% of responses), we report complete results only for GPT-3.5 and GPT-4o.

\begin{table}[t]
    \centering
    \caption{\textbf{Performance Comparison: Prompted LLMs vs. \TheName{}} Metrics for CWE classification and binary vulnerability prediction.}
    \label{tab:rq1_kastro_vs_prompted_llms}
    \resizebox{0.9\columnwidth}{!}{
    \begin{tabular}{l|l|c|c|c}
        \toprule
        \textbf{Dataset} & \textbf{Metric} & \textbf{GPT-3.5} & \textbf{GPT-4o} & \textbf{\TheName{}} \\
        \midrule
        \multirow{3}{*}{BigVul} 
        & F1 & 7.19 & 9.83 & \textbf{76.31} \\
        & Precision & 8.76 & 10.51 & \textbf{76.92} \\
        & Recall & 19.56 & 18.70 & \textbf{76.47} \\
        \midrule
        \multirow{3}{*}{BigVul-bin} 
        & F1 & 7.38 & 9.39 & \textbf{47.19} \\
        & Precision & 6.05 & 5.25 & \textbf{59.17} \\
        & Recall & 9.44 & \textbf{44.26} & 39.24 \\
        \bottomrule
    \end{tabular}}
\end{table}

\textbf{Key Findings.} \TheName{} significantly outperforms GPT-3.5 and GPT-4o across all metrics. On BigVul, \TheName{} achieves a weighted F1 score of 76.31 compared to 7.19 (GPT-3.5) and 9.83 (GPT-4o). Similarly, for binary vulnerability classification (BigVul-bin), \TheName{} achieves an F1 of 47.19, surpassing the performance of GPT-3.5 (7.38) and GPT-4o (9.39). These results underscore the limitations of generic LLMs in vulnerability classification tasks.

\textbf{Limitations of LLMs.} While the LLMs provide reasoning for their predictions, their inability to handle complex CWE-specific classifications highlights the necessity for specialized models like \TheName{}. Detailed reasoning analysis for LLM predictions remains an avenue for future exploration.

\subsection{\textbf{RQ1: Other LLMs vs. \TheName{}}}
\label{sec:rq1}

\textbf{Objective.} This research question evaluates how \TheName{} performs in comparison to state-of-the-art LLMs (e.g., GPT-3.5 and GPT-4o) on CWE classification and binary vulnerability prediction tasks across multiple datasets.

\textbf{Model Variants.} \TheName{} was tested with two configuration options:
\begin{itemize}
    \item \textbf{with-mask}: Enforces masking in the learned attention matrix, ensuring the model focuses on connected nodes in augmented ASTs. After experimentation, we fixed \textbf{with-mask} to True.
    \item \textbf{embedding-type}: Two embedding models were evaluated, OpenAI's \footnote{\texttt{text-embedding-3-small}} (``small'') and \footnote{\texttt{text-embedding-ada-002}} (``ada''). The ``small'' embedding model demonstrated superior performance and was selected for all experiments.
\end{itemize}

\begin{table}[t]
    \centering
    \small
    \caption{\textbf{\TheName{} Model Performance.} Weighted metrics across datasets. BigVul: 36 classes, BigVul-bin: 2 classes, DiverseVul: 49 classes, PrimeVul: 4 classes. PrimeVul's test set contains only $\approx$3\% vulnerable functions, hence binary results are omitted.}
    \label{tab:kastro-performance}
    \resizebox{0.9\columnwidth}{!}{%
    \begin{tabular}{l|c|c|c|c}
        \toprule
        \textbf{Dataset} & \textbf{F1} & \textbf{Precision} & \textbf{Recall} & \textbf{Accuracy} \\
        \midrule
        BigVul~\cite{bigvul} & 76.31 & 76.92 & 76.47 & 76.47 \\
        BigVul-bin & 47.19 & 59.17 & 39.24 & 97.28 \\
        DiverseVul~\cite{diversevul} & 69.12 & 69.64 & 69.14 & 69.14 \\
        PrimeVul~\cite{primevul} & 92.33 & 90.99 & 93.78 & 93.78 \\
        \bottomrule
    \end{tabular}}
\end{table}

\textbf{Findings.} \TheName{} consistently outperforms GPT-3.5 and GPT-4o on all datasets (Table~\ref{tab:kastro-performance}). For instance:
\begin{itemize}
    \item On BigVul, GPT-3.5 achieves a weighted F1 of 7.19 for CWE classification, and GPT-4o scores 9.83, while \TheName{} achieves 76.31.
    \item On BigVul-bin, \TheName{} achieves a weighted F1 of 47.19 compared to 7.38 (GPT-3.5) and 9.39 (GPT-4o).
\end{itemize}

\textbf{Comparison with Larger Models.} Despite having only 1M parameters, \TheName{} achieves competitive results compared to significantly larger models like GraphCodeBERT~\cite{graphcodebert} (125M), PolyCoder~\cite{polycoder} (2.7B), and T5~\cite{codet5} (220M), none of which exceed 50\% F1 on the DiverseVul dataset. This suggests that specialized small models can outperform general-purpose large models for domain-specific tasks.

\subsection{\textbf{RQ2: CWE-Specific Performance}}
\label{sec:rq2}

The results from \TheName{} inference on the test sets of BigVul and DiverseVul provide a comprehensive evaluation of its ability to handle diverse CWE classes. To assess its performance, we conducted experiments focusing on class-specific metrics, including Precision, Recall, and F1 scores, averaged across individual classes for each dataset.

\textbf{Experimental Setup.} For this evaluation, \TheName{} was tasked with classifying functions based on their associated CWE labels in the BigVul and DiverseVul test sets. These datasets include a mix of specific CWEs and broader categories, offering a diverse challenge. By analyzing the class-wise metrics, we aimed to identify the strengths and weaknesses of the model across different types of CWEs.

\textbf{Performance Consistency on BigVul.} On BigVul, \TheName{} exhibits consistent performance across the 36 CWE classes, achieving an average F1 score of approximately 0.8. The model performs particularly well on CWE-269: "Improper Privilege Management" and CWE-704: "Incorrect Type Conversion or Cast," achieving F1 scores above 0.9. These results highlight the model's capability to identify specific and well-defined vulnerabilities. However, the model struggles with broader categories like CWE-388: "7PK - Errors," which aggregate multiple specific CWEs, making classification inherently more challenging. This discrepancy is likely due to the lack of granularity in such categories, which can obscure the patterns needed for accurate classification.

\textbf{Variability in DiverseVul Results.} In contrast, results on DiverseVul exhibit greater variability across its 49 CWE classes. CWE-212: "Improper Removal of Sensitive Information Before Storage or Transfer" achieves an almost perfect F1 score, demonstrating the model's ability to handle specific and clearly defined vulnerabilities. Other strong performers include CWE-191: "Integer Underflow" and CWE-613: "Insufficient Session Expiration." However, some classes, such as CWE-122: "Heap-based Buffer Overflow" and CWE-19: "Data Processing Errors," exhibit lower F1 scores, highlighting areas where the model's performance is limited. Notably, many of these underperforming classes are categorized as general groupings, which further complicates classification.

\subsection{\textbf{RQ3: Model Efficiency}}
\label{sec:rq3}

We assess \TheName{}'s efficiency by evaluating its inference performance across different datasets using GPU and CPU setups. Despite its lightweight architecture, \TheName{} demonstrates excellent scalability and speed, making it well-suited for processing large datasets.

\TheName{} is a compact Transformer model tailored for binary vulnerability prediction and CWE classification. It includes a single encoder layer with multi-head attention that integrates $K$ augmented AST interaction matrices. With approximately 1 million trainable parameters, \TheName{} occupies only $\approx$4MB of disk space when trained, showcasing remarkable storage efficiency.

In our experiments, \TheName{} was trained for 5 rounds of 5 epochs, each with distinct random seeds for reproducibility. Training completes within 1-3 hours, depending on dataset size, and inference takes just seconds per full pass. This lightweight design ensures minimal overhead while maintaining high performance.

Table~\ref{tab:rq3_kastro_efficiency} summarizes the evaluation metrics across datasets. On the GPU, \TheName{} achieves an impressive throughput of 5,434 samples per second for BigVul-bin, while on the CPU, it processes up to 12,989 samples per second for the same dataset. Across all datasets, throughput consistently exceeds 1,500 samples per second on both GPU and CPU setups.

\textbf{Compact and Efficient Design.} These results highlight \TheName{}'s suitability for practical deployment in resource-constrained environments. Its compact size and rapid inference capabilities make it an ideal choice for real-world applications requiring efficient and accurate vulnerability detection.

\begin{table*}[t]
    \centering
    \caption{\textbf{\TheName{} Inference Efficiency.} Inference throughput and evaluation time for different datasets on GPU and CPU. The trained model is compact ($\approx$4MB) and achieves competitive inference speeds, processing thousands of samples per second.}
    \label{tab:rq3_kastro_efficiency}
    \resizebox{0.9\textwidth}{!}{%
    \begin{tabular}{l|c|c|c|c|c}
        \toprule
        \textbf{Dataset} & \textbf{\# Samples} & \textbf{Eval Time GPU (s)} & \textbf{Eval Time CPU (s)} & \textbf{GPU Throughput (samples/s)} & \textbf{CPU Throughput (samples/s)} \\
        \midrule
        BigVul & 31,894 & 20.52 & 20.64 & 1,554 & 1,545 \\
        BigVul-bin & 31,953 & 5.88 & 2.46 & 5,434 & 12,989 \\
        DiverseVul & 24,601 & 14.65 & 17.07 & 1,679 & 1,441 \\
        PrimeVul & 24,990 & 15.03 & 15.91 & 1,662 & 1,570 \\
        \bottomrule
    \end{tabular}%
    }
\end{table*}

\section{Related Work}\label{sec:related}

This section reviews previous studies in three primary areas: ML-driven vulnerability detection, the application of LLMs in security, and strategies for LLM adaptation to specific tasks.

\subsection{ML-Driven Vulnerability Detection}
Machine learning approaches have become increasingly desirable for vulnerability detection as they require less manual effort compared to pattern-based techniques like FlawFinder \cite{ferschke2012flawfinder} and ITS4 \cite{ITS4-2000-viega}. Early ML-based methods, such as VulDeePecker \cite{Li_2018}, utilized Bi-LSTMs with word2vec encodings of code gadgets, demonstrating significant improvements over traditional rule-based systems.

Abstract syntax tree (AST)-based methods have also gained traction. \cite{lin-2017-vulnerability} explored AST serialization combined with Bi-LSTMs for function-level vulnerability detection. Building on these, \cite{enhancing-data-augmentation} employed neural sub-tree encodings to capture fine-grained syntactic features, while \cite{ast_vuln_extrap} extended these approaches for generalized vulnerability extrapolation using AST-based representations.

Recently, transformer-based models have dominated vulnerability detection tasks. \cite{thapa_asac22} demonstrated that fine-tuned models like CodeBERT \cite{feng2020codebert} and GraphCodeBERT \cite{graphcodebert} significantly outperform Bi-LSTMs. Further, \cite{vulnerability-execution-paths} emphasized learning from syntax-based execution paths to enhance detection performance. Despite these advances, existing methods often fall short in leveraging augmented AST structures to address generalization and interpretability, a gap addressed by \TheName{}.

\subsection{LLMs in Computer Security}
Large Language Models (LLMs) have seen growing use in security applications, including vulnerability repair \cite{fixing-security-bugs-ahmad-2023, pearce-vul-repair-2023}, CWE mapping \cite{liu-etal-2023-end}, and policy analysis \cite{divas-paria-2023}. Despite their promise, off-the-shelf LLMs like GPT-3.5 and GPT-4o have shown limited success in vulnerability detection \cite{chatgpt-eval-vd-cheshkov-2023, chatgpt-for-vd-2023-fu}.

Recent frameworks have sought to improve LLM-driven detection through tailored strategies. For instance, \cite{DLAP-framework} introduced deep-learning-augmented prompting frameworks, while \cite{enhancing-data-augmentation} utilized vulnerability-preserving data augmentation to enrich training data. However, challenges remain in adapting these general-purpose models for domain-specific tasks. Our work addresses this by integrating AST-based structural insights with lightweight fine-tuning techniques, enabling robust performance even with limited computational resources.

\subsection{LLM Adaptation}
LLM adaptation focuses on resource-efficient strategies for task-specific fine-tuning. Techniques like prompt tuning \cite{lester2021power} and prefix tuning \cite{li-liang-2021-prefix} enable parameter-efficient updates by optimizing input embeddings. Similarly, LoRA \cite{hu2022lora} employs low-rank approximations to reduce the number of trainable parameters.

Adapter-based methods, such as \cite{houlsby2019parameterefficient} and \cite{hu2023llmadapters}, introduce small auxiliary modules between transformer layers to achieve efficient fine-tuning. These methods have been widely adopted for various NLP and security tasks. In contrast, our approach avoids auxiliary modules, opting instead for a single lightweight transformer block with multi-head attention to integrate AST-derived biases into the final prediction.

By combining text embeddings with augmented AST structures, \TheName{} achieves strong performance on vulnerability detection tasks while maintaining computational efficiency. This unique adaptation mechanism sets our approach apart from existing parameter-efficient methods and bridges the gap between LLM generalization and domain-specific performance.

\section{Discussion and Future Work}

\subsection{Discussion}

Our experimental results demonstrate a key finding: a lightweight, specialized model like \TheName{} can significantly outperform massive, general-purpose LLMs on the nuanced task of vulnerability detection. This success is not merely a matter of parameter count but of architectural design. By unifying high-level semantic information from LLM embeddings with explicit structural knowledge from Abstract Syntax Trees (ASTs), \TheName{} incorporates a crucial inductive bias for code structure. Its structure-aware attention mechanism guides the model to focus on syntactic relationships that are often indicative of vulnerabilities—a targeted approach that stands in contrast to the more generalized, and often less effective, pattern matching of prompted LLMs.

The practical implications of this approach are significant. With only 1 million parameters and a disk footprint of about 4MB, \TheName{} is highly efficient. It trains within hours and executes inference in seconds on standard CPUs, making it ideal for real-world, resource-constrained environments such as local IDE plugins or CI/CD pipeline scanners. This democratizes access to advanced security analysis, enabling private and low-cost code scanning without reliance on expensive GPU hardware. Furthermore, our analysis of CWE-specific performance shows that the model excels at identifying vulnerabilities with clear, consistent structural patterns (e.g., CWE-269 "Improper Privilege Management") while struggling with broader, more abstract categories (e.g., CWE-388 "7PK - Errors"). This suggests that the granularity of vulnerability labels directly impacts model performance.

\subsection{Future Work}

Based on the foundation of \TheName{}, we propose several directions for future research to broaden its impact and capabilities:

\begin{itemize}
    \item \textbf{Inter-Procedural and Cross-File Analysis:} A primary limitation of the current approach is its focus on single-function analysis without leveraging surrounding code context. Many vulnerabilities, however, arise from complex interactions between different functions, potentially across multiple files. We plan to extend \TheName{} to perform inter-procedural analysis by constructing program-level call graphs or data-flow graphs, using the model's function representations as inputs to a graph neural network (GNN) to detect more complex, inter-connected vulnerabilities.

    \item \textbf{Enhanced Interpretability and Automated Repair:} To move beyond mere detection, we aim to improve the actionability of the model's findings. The structure-aware attention mechanism provides a natural path toward explainability (XAI); by visualizing the attention weights, we can highlight the specific code constructs and AST subtrees that contribute most to a vulnerability prediction. These insights could, in turn, be used to guide a generative model to suggest targeted and syntactically correct code patches, moving closer to a system for automated vulnerability repair.

    \item \textbf{Expanding Language Support and Embedding Models:} While this work focuses on C/C++ due to data availability, the core architecture is language-agnostic. A crucial next step is to adapt the framework for other languages like Java, Python, and JavaScript by integrating their respective AST parsers. Concurrently, we will explore replacing the proprietary OpenAI embeddings with open-source, code-specialized models (e.g., CodeT5, GraphCodeBERT) to create a fully self-contained and potentially more powerful vulnerability detection tool.
\end{itemize}

\section{Conclusion}
\label{sec:conclusion}

We present \TheName{}, a lightweight Transformer model tailored for few-shot vulnerability detection in C/C++ source code. By integrating AST-based data augmentation and sparse attention with text embeddings, \TheName{} harmonizes structural and semantic features to enhance vulnerability detection and CWE classification. Unlike off-the-shelf LLMs, which struggle with this task due to biases towards non-vulnerable code, \TheName{} leverages its compact architecture to achieve competitive performance with state-of-the-art models while maintaining efficiency and simplicity.

With $\approx$1M parameters occupying only 4MB on disk, \TheName{} trains within hours and performs rapid inference on CPUs, enabling secure, local processing of sensitive code without reliance on resource-intensive GPU servers. Our results demonstrate that \TheName{} matches or exceeds the performance of larger industrial LLMs, underscoring the value of combining structural and semantic insights for vulnerability detection. To encourage further research, we release all software artifacts to the community.

\subsection{Limitations}
\label{sec:limitations}

\TheName{} focuses exclusively on C/C++ source code, chosen for its prevalence, available data, and the complexity of vulnerabilities in these languages. While the approach is conceptually applicable to other programming languages, dataset limitations and embedding constraints shaped our experiments. Specifically, our input length is limited by the token capacity of embedding models, resulting in the omission of a small subset of functions. Moreover, we do not explore variations in embedding dimensionality or broader dataset refinements, which could provide opportunities for future work. Despite these constraints, \TheName{} demonstrates robust performance and offers a promising foundation for lightweight vulnerability prediction.

\bibliographystyle{ACM-Reference-Format}
\bibliography{acmart}

\end{document}